\documentclass[debug,overfull]{epl}
\usepackage{amssymb}

\shortauthor{J.P. Wittmer \etal}
\shorttitle{Vibrations of nanometric structures}
\institute{
\inst{1}
D\'epartement de Physique des Mat\'eriaux, CNRS,
Universit\'e de Lyon, 43 Boulevard du 11 Novembre, 69622
Villeurbanne Cedex, France.
\inst{2}
D\'epartement de Physique, Universit\'e de Montr\'eal,
C.P. 6128, Succ. Centre-Ville, Montr\'eal, Québec, CANADA H3C 3J7.
}
\pacs{72.80.Ng}{Disordered solids}
\pacs{65.60.+a}{Thermal properties of amorphous solids and glasses}
\pacs{61.46.+w}{Nanoscale materials: nanoparticles}
\rec{}{}
\begin{document}

\title{Vibrations of amorphous, nanometric structures: When does continuum theory apply?}
\author{J.~P.~Wittmer\inst{1}\thanks{E-mail: jwittmer@dpm.univ-lyon1.fr}
\and A.~Tanguy\inst{1} \and J.-L.~Barrat\inst{1} \and L.~Lewis\inst{2}}
\maketitle
%

\begin{abstract}
We investigate the low frequency end of the vibrational spectrum
in small (nanometric) disordered systems. Using numerical
simulation and exact diagonalization for simple two dimensional
models, we show that continuum elasticity, applied to these
systems, actually breaks down below a length scale of typically 30
to 50 molecular sizes. This length scale is likely related to the
one which is generally invoked to explain the peculiar vibrational
properties of glassy systems.
\end{abstract}


Determining the vibrational properties
--- i.e. the vibration
frequencies and the associated displacement fields ---
of solid bodies with various shapes is a
well studied area of continuum mechanics.
The early works of Lamb or Rayleigh \cite{elastodynamics},
who determined these
vibrations using classical elasticity theory of isotropic materials,
have found applications in fields
as different as planetary science and nuclear physics. In today's
materials sciences, the increasing development of materials
containing nanometer size structures naturally leads one to
question the limits of applicability of the continuum elasticity
theory, which is in principle valid only on length scales much
larger than the interatomic distances \cite{landau,alexander}.
Investigating the vibration modes of nanometric objects using
atomic level simulations is a natural way of probing this
applicability. Such an investigation is particularly relevant from
an experimental viewpoint, since these  properties, inferred from
spectroscopic measurements, are systematically interpreted within
the framework of continuum elasticity \cite{vallee,hodak,saviot}.
In this paper, we use a simple model, which is commonly used  as a
generic model for  liquids and amorphous \cite{cryst} solids, 
to investigate the applicability of continuum theory at small scales.

In a computer simulated system, in which all particle coordinates
and interparticle forces are exactly known, it is possible to calculate
exactly the vibration frequencies around an equilibrium position.
This is achieved by exact diagonalization
of the so called dynamical matrix \cite{Kittel}, a $(d N)\times (d N)$
(where $d$ is the number of spatial dimensions and $N$ the number of particles)
matrix expressible in terms of the first and second derivatives of the interparticle
interaction potentials.
The corresponding displacement fields are given by the
eigenvectors of the dynamical matrix. We have carried out a
systematic comparison of the exact eigenfrequencies calculated in
this way with those predicted by continuum elasticity, for two
dimensional (d=2) objects of increasingly large sizes. The objects
we consider are either disk shaped clusters of diameter  $2R$, or
bulk like systems contained in a square of side $L$ with periodic
boundary conditions. Microscopically, they are formed by
quenching a slightly polydisperse liquid of spherical particles
interacting via simple Lennard-Jones pair potentials into the
nearest energy minimum. For the clusters, the quench is realized
after cutting a disk out of a much larger bulk like sample. The
resulting structures are amorphous, i.e. exhibit no crystal like order.

\begin{figure}
\onefigure[scale=0.5]{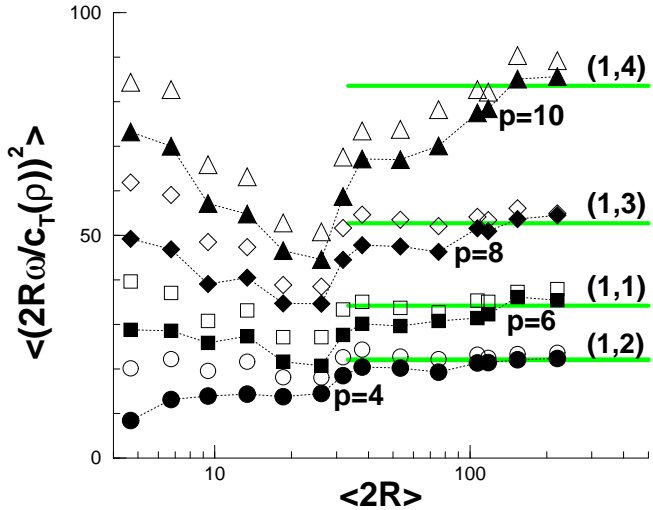}
\caption{Eigenfrequencies for
disk shaped clusters. Squares of the eigenfrequencies, versus
diameter $2R$, for the lowest frequency vibration modes of disk
shape clusters of increasing sizes. Each symbol corresponds to a
different mode number $p$. The value of  $p$ is indicated for each
series of black symbols (even values of $p$). The first nontrivial
mode corresponds to $p=4$, since the first three modes correspond
to global translations and rotation. The frequencies have been
multiplied by the size and by the sound velocity, so that in the
limit of large systems they should regroup into degenerate pairs
and tend to a constant value. These quantized values can be
calculated explicitly from the theory of elasticity (horizontal
lines). The quantum numbers $(k,n)$ associated to each of these
eigenmodes within elastic theory are also indicated.} \label{fig1}
\end{figure}

We have generated such two dimensional amorphous clusters with
sizes ranging from $R=3\sigma $ to $R=109\sigma$, where $\sigma$
is the particle diameter. This corresponds to numbers of atoms
between $N=16$ and $N=32768$. A comparable range has been studied
for the bulk like materials, with $ 7 <L/\sigma < 208$ and $
50<N<40000$. The density of the bulk like materials corresponds to
a near zero pressure state, while that of the clusters slightly depends
on their radius. A smaller radius implies a higher capillary
pressure, therefore a slightly higher density. For each system,
the lowest (100 first) vibration eigenfrequencies and eigenvectors
have been determined using the version of the Lanczos method
implemented in the PARPACK numerical package \cite{arpack}. We
concentrate on the lowest end of the vibrational spectrum, since
this is the part that corresponds to the largest wavelengths for
the vibrations. Hence, the corresponding modes are those for which
one would expect continuum theory to be applicable. They are also
those which are probed in low frequency Raman scattering
experiments \cite{saviot}, in order to determine the typical size
of nanoparticles.

\begin{figure}
\onefigure[scale=0.5]{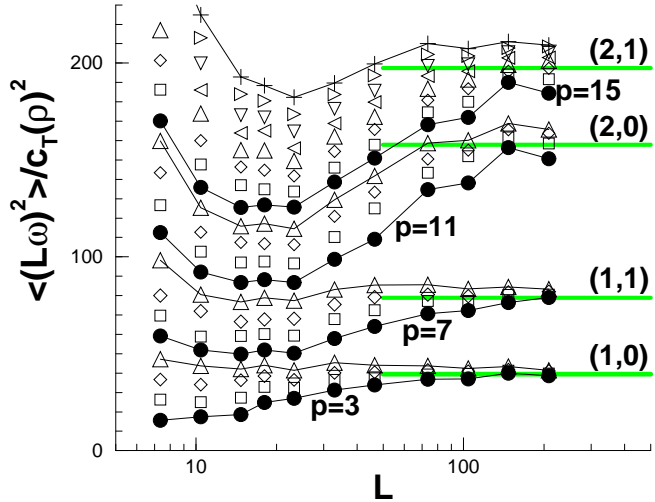}
\caption{
Same as figure~\protect\ref{fig1}, for bulk like systems with periodic
boundary conditions in a square box of lateral size $L$. Again, a series
of identical symbols corresponds to a given mode number $p$, starting
at $p=3$ (2 global translations have zero frequency).
In the elastic limit, the modes
are now quantized by two integers $(n,m)$. Lowest frequency  modes,
having either $n=m$, $n=0$ or $m=0$, have a fourfold degeneracy.
The highest frequency mode ($n=1,m=2$) has eightfold degeneracy.}
\label{fig2}
\end{figure}

Once the exact modes are determined, they may be compared to those
obtained from the continuum elasticity theory. For bulk like
systems with periodic boundary conditions, these elastic  modes
are simply plane waves, and the eigenfrequencies are quantized by
two integers $(n,m)$ that define the periodicities in the $x$ and
$y$ directions, respectively.  The plane waves may be transverse,
with a displacement field orthogonal to the direction of
propagation, or longitudinal, with the displacement parallel to
this direction. Quantitatively, the frequencies of the transverse
waves are of the form $\omega_{nm}= {2\pi \over L} c_T
\sqrt{n^2+m^2}$, where $c_T$ is the transverse sound velocity. A
similar formula is obtained for longitudinal waves, with a
velocity $c_L>c_T$. Hence the eigenfrequency associated with a
pair of different integers $n,m$ has a four-fold (or eight-fold if
$n\neq m\neq 0$) degeneracy, associated with waves travelling in
two opposite and orthogonal directions. The longitudinal and
transverse sound velocities can be expressed in terms of the
Lam\'e coefficients $\lambda$ and $\mu$   \cite{landau}, $c_T=
\sqrt{\mu/\rho}$ and $c_L=\sqrt{(\lambda+2\mu)/\rho}$, where $\rho$
is the mass density.

The situation for disk-shaped objects is somewhat more complex,
with again each vibration characterized by two quantum numbers $n$
and $k$, and eigen-frequencies of the form $\omega_{n}={2k\pi\over
R }c_T f_{nk}(\nu)$.  Here $\nu$ is the Poisson ratio of the
material. The quantum number $n$ is associated with the angular
dependency of the displacement field, and the number $k$ to its
radial dependency. For $n>0$, each mode exhibits a twofold
degeneracy, corresponding to oscillations in two orthogonal
directions. Degenerate eigenvalues are therefore inherent to the
continuum treatment of systems which are assumed to be highly
symmetric.

\begin{figure}[ht]
\onefigure[scale=0.65]{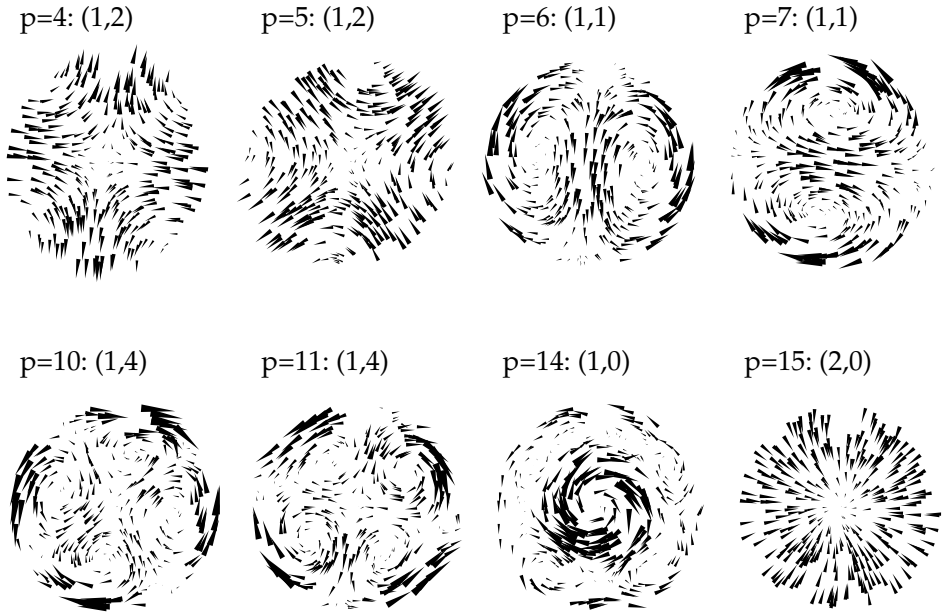}
\vspace*{-4.7cm} 
\caption{
Schematic representation of the displacement fields associated to
some low frequency modes in a large, disk-shaped cluster with
10000 particles ($2R=120$). For such large clusters, the
displacement fields essentially conform to standard elastic
behavior. The two quantum numbers $k$ and $n$ are indicated for
each mode. The twofold degeneracy, corresponding to displacements
fields rotated by 90 degrees, is illustrated in the first six
fields. The last two correspond to modes with $n=0$, and their
frequency is not degenerated. Note that the last mode represented
is a 'breathing' mode, with a relatively high frequency (the 15th
by increasing magnitudes).} \label{fig3}
\end{figure}

Figures~\ref{fig1} and ~\ref{fig2} present the results of the comparison
between elastic theory predictions
and exact diagonalization for the eigenvalues, for systems of variable sizes.
Elastic theory would predict that $\omega L$ ($2\omega R$ for clusters)
values are independent of the size, and highly degenerated. The
figures show that this characteristic degeneracy is recovered
only for large enough sizes, typically $L$ or $2R>40\sigma$.

The elastic limits are obtained from the analytic formulae described above,
by inserting the appropriate values of the Lam\'e coefficients.
These values in turn are determined  numerically, from the changes
in energy associated with a small compression (or shear
deformation) of a large, bulk sample. The structure of the
displacement fields for some of the lowest eigenmodes  of a large
cluster ($R=60$) is shown in figure~\ref{fig3}.

Our results clearly indicate the existence of a typical size,
$L\simeq 40\sigma$, below which the predictions of continuum elasticity become
erroneous. The classical degeneracy of the vibration
eigen-frequencies is lifted, and the resulting density of states,
when averaged over many realizations (typically 20), becomes a
continuous function. Figures~\ref{fig1} and ~\ref{fig2}
show that the approach of the
elastic limit is nonmonotonous, with systems of intermediate sizes
having eigenfrequencies lower than predicted by the elastic
theory, while very small systems vibrate at higher frequencies.

It is therefore interesting to speculate on the nature of the
microscopic features that become relevant  on this length scale,
and make the elastic approach inappropriate. Density variations
are easily excluded, since the density is always found to be
perfectly homogeneous down to a scale typical of the interatomic
separation.  An interesting alternative is to look for
inhomogeneities in the local stresses at the atomic scale, which
could lead to local anisotropies responsible for the lifting the
elastic degeneracy. Note that although the system is lying in an
energy minimum, its disordered character implies that the forces
between pairs of neighboring particles are, as a rule, nonzero, so
that local stresses are frozen into each sample.

\begin{figure}
\onefigure[scale=0.4]{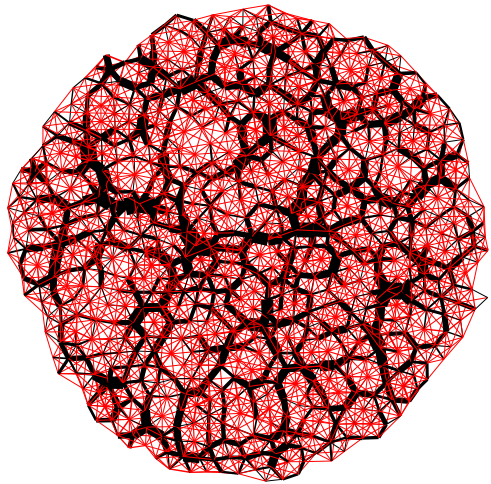}
\vspace*{0.3cm}
\caption{
Representation of the force network frozen in a large,
amorphous cluster ($2R \approx 30$).
Atoms lie at the vertices, and bonds represent the interaction forces.
Attractive forces are represented by red bonds, and repulsive ones by black bonds.
The thickness of each bond is proportional to the magnitude of the associated
force.}
\label{fig4}
\end{figure}

A pictorial representation of these stresses is given in figure~\ref{fig4},
which shows the network of repulsive (black) and attractive (red) forces
in a large, disk-shaped cluster. These stresses are widely distributed.
Visual inspection reveals a characteristic length scale
of the repulsive forces network, much larger than the interatomic separation.
Interestingly, the figure also reveals the existence of 'force
chains' very similar to those found in granular media under stress
\cite{forcechains}.  This feature may be added to the list of
similarities which have been noticed between granular and
amorphous (glassy) materials \cite{liunagel}. Note, however, that
the existence of an elastic limit for large systems, which is
clearly established here for amorphous systems, is still a matter
of debate in granular systems \cite{elasticgranular}.

\begin{figure}
\onefigure[scale=0.4]{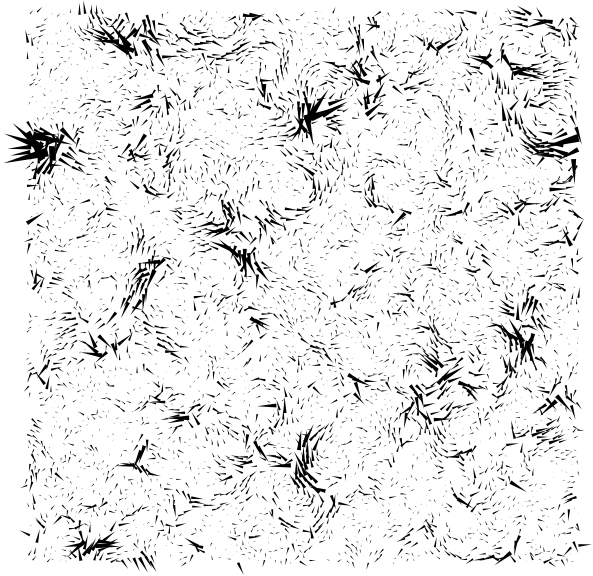}
\caption{Non-affine component of the displacement field.
A large sample
(10000 particles, $L=104$) with periodic boundary conditions has been
submitted to an asymptotically weak elongational strain in horizontal direction.
The resulting displacement
field is computed after relaxing the atoms to their new
equilibrium positions.
The vector field represented on the figure is
(proportional to) the {\em non-affine} part of this displacement field,
i.e. the displacement from which the affine component proportional to the
homogeneous strain has been substracted.
}
\label{fig5}
\end{figure}

Another way of illustrating the failure of classical elasticity at
small scales is to investigate the displacement field in a large,
deformed sample. If the system is uniformly strained at large
scales (e.g. by compressing a rectangular box), classical
elasticity implies that  the strain is uniform at all scales, so
that the atomic displacement field should be affine with respect
to the box deformation. We have investigated the nonaffine
component of the atomic displacement field in large systems
subjected to a compression, and found that the 'elastic length'
discussed above manifests itself through {\em correlated
deviations} from a purely affine displacement. In some regions,
the displacement is much larger than expected from a purely affine
transformation, as illustrated in figure~\ref{fig5}. The size of these
regions is similar, albeit somewhat smaller, than the one at which
the results of continuum elasticity for the vibration modes are recovered.

An important consequence of the existence of these "softer"
regions is that  the  Lam\'e coefficients have to be computed from
the change in  total  energy due to the actual elastic
displacement field  following a deformation (an elongation in the
case of figure \ref{fig5}) . These coefficients can differ, by a
factor as large as $2$, from those that would result from an
homogeneous deformation. The latter are obtained from standard
formulae for elastic constants \cite{hoover} such as
\cite{hoover1}
\begin{equation}
\mu = \frac{1}{V} \sum_{ij} \left(\frac{(\phi^{\prime \prime}
(r_{ij})}{r_{ij}^2}- \frac{\phi^{\prime} (r_{ij})}{r_{ij}^3}\right)
(x_i-x_j)^2 (y_i-y_j)^2.
 \label{EC}
\end{equation}
The  derivation of equation \ref{EC}, however, implicitly assumes
that the deformations are affine on all scales. Using such a
formula, one indeed  obtains a value for $\mu$ which is much too
large to account for our results on vibrational frequencies.
Clearly, a calculation taking into account the
nonaffine character of the displacements is necessary
for disordered systems. For such a calculation,
a system size larger than the
correlation length of the non-affine displacement field is needed
to obtain a reliable estimate of the Lam\'e coefficients. Below
this length scale, the local elastic coefficients are
inhomogeneous. Inhomogeneities in local elastic coefficients,
which are explicitly displayed in figure \ref{fig5}, are an
essential ingredient in several recent calculations on disordered
elastic systems \cite{Schirmacher,Ruocco}. Our work provides a
first estimate for the typical length scale of these
inhomogeneities.

Using computer generated amorphous structures, we have thus shown
that the application of continuum elasticity theory is subject
to strong limitations in amorphous solids, below a length scale of
typically $40$ interatomic distances.  This length is surprisingly
large, considering the short range of the interatomic
correlations.  The origin of the departure from elastic behavior
is very likely related to the disorder in interatomic interactions
(local stresses and force constants). This disorder is illustrated
by the structure of the force network frozen into the solid, as
shown in figure~\ref{fig4}.
Local inhomogeneities in elastic constants are
revealed by the investigation of the nonaffine part of the
displacement field when a large sample is deformed, and reveal a
similar length scale.
Interestingly, similar, or somewhat smaller,
sizes are often invoked \cite{glassyvibs1,glassyvibs2}, as typical
of the heterogeneities that give rise to anomalies in the
vibrational properties of disordered solids (glasses) in the
Terahertz frequency domain, the so-called boson peak.
In particular, reference \cite{glassyvibs2} considers the
existence of rigid domains separated by softer interfacial zones,
not unlike those revealed by the nonaffine displacement pattern of
figure~\ref{fig5}. Our work offers a new point of view on this
feature. At the wavelength corresponding to these Thz vibrations,
comparing the vibrational density of states to that of a
continuum, elastic model (the Debye model) is not necessarily meaningful. 
It also suggests that numerical investigation of
vibrational properties in disordered systems should systematically
make use of large samples ($L\gg 40\sigma$), in order to
avoid finite-size effects  and to include correctly 
contributions from ``softer" elastic regions. 
Although we are aware that this
conclusion may have to be modified in a three dimensional space,
careful finite-size studies should also be undertaken in that case
\cite{note1}.


\begin{thebibliography}{99}
\bibitem{cryst}
Depending on the sample preparation method, nanosized particles may be in fact 
more or less crystalline, rather than amorphous. However, as amorphous clusters
of imperfect crystallites are possible the query raised in the title is relevant.
\bibitem{hoover1} In this expression, $\phi$ is the interparticle
pair potential, $\phi^{\prime}$ and $\phi^{\prime \prime}$ its
derivatives, and the sum runs over all pairs of particles. The
formula has been adapted from ref. \cite{hoover} for the special
case of zero temperature.
\bibitem{note1} Indeed, this study initially originated from an attempt
to compute the vibrational modes of three dimensional clusters. In
three dimensions, we were unable to reach the elastic limit for
the larger sizes (10000 atoms) that could be investigated at the time.
\end{thebibliography}
\end{document}